\documentclass[
	aps, pra, superscriptaddress, twocolumn,
	10pt
	floatfix, 
    nofootinbib,
	tightenlines
]{revtex4-1}

\usepackage[final]{graphicx}
\usepackage{times,bbm,amsmath,amssymb}
\usepackage{epsfig,color}
\usepackage{xcolor}
\usepackage{hyperref}
\hypersetup{
    colorlinks = true
}
\usepackage{cleveref}
\usepackage{microtype}

\usepackage{float,siunitx}
\usepackage[caption = false]{subfig}

\usepackage[greek,english]{babel}
\usepackage{thumbpdf,enumerate}
\usepackage{booktabs}
\usepackage{sidecap}
\usepackage[scaled=.8]{couriers}
\usepackage{multirow}
\usepackage{placeins}
\usepackage{relsize}
\usepackage{pst-grad,bm}
\usepackage{epigraph}\usepackage{epigraph}

\usepackage{gensymb}
\usepackage{longtable}
\usepackage{ulem} 
\usepackage{svg}
\usepackage{acronym}
\usepackage{physics}
\DeclareUnicodeCharacter{0301}{\'{e}}

\begin{document}

%\author{Alessia Suprano}
%\author{Danilo Zia}

\address{Dipartimento di Fisica, Sapienza Universit\`{a} di Roma, Piazzale Aldo Moro 5, I-00185 Roma, Italy}

%\author{Fabio Sciarrino }

%\email{fabio.sciarrino@uniroma1.it}

%\address{Dipartimento di Fisica, Sapienza Universit\`{a} di Roma, Piazzale Aldo Moro 5, I-00185 Roma, Italy}

\vspace{10pt}

%\title{High dimensional \textit{intra}- and \textit{inter}- particle entangled states generated via a quantum dot source} 

\title{Orbital angular momentum based \textit{intra}- and \textit{inter}- particle entangled states generated via a quantum dot source}

\author{Alessia Suprano}

\author{Danilo Zia}
\address{Dipartimento di Fisica, Sapienza Universit\`{a} di Roma, Piazzale Aldo Moro 5, I-00185 Roma, Italy}

\author{ Mathias Pont} 
\address{Centre for Nanosciences and Nanotechnology, CNRS, Universit\'e Paris-Saclay,
UMR 9001,10 Boulevard Thomas Gobert, 91120, Palaiseau, France}

\author{Taira Giordani}

\author{Giovanni Rodari }
\address{Dipartimento di Fisica, Sapienza Universit\`{a} di Roma, Piazzale Aldo Moro 5, I-00185 Roma, Italy}

\author{Mauro Valeri}
\address{Dipartimento di Fisica, Sapienza Universit\`{a} di Roma, Piazzale Aldo Moro 5, I-00185 Roma, Italy}

\author{Bruno Piccirillo }
\address{Dipartimento di Fisica "Ettore Pancini", Universit\`{a} di Napoli Federico II, Complesso Universitario di Monte Sant'Angelo, Via Cintia, 80126 Napoli, Italy}
\address{INFN – Sezione di Napoli, Via Cintia, 80126 Napoli, Italy}

\author{Gonzalo Carvacho }
\address{Dipartimento di Fisica, Sapienza Universit\`{a} di Roma, Piazzale Aldo Moro 5, I-00185 Roma, Italy}

\author{Nicolò Spagnolo }
\address{Dipartimento di Fisica, Sapienza Universit\`{a} di Roma, Piazzale Aldo Moro 5, I-00185 Roma, Italy}

\author{Pascale Senellart}
\address{Centre for Nanosciences and Nanotechnology, CNRS, Universit\'e Paris-Saclay,
UMR 9001,10 Boulevard Thomas Gobert, 91120, Palaiseau, France}

\author{Lorenzo Marrucci}
\address{Dipartimento di Fisica "Ettore Pancini", Universit\`{a} di Napoli Federico II, Complesso Universitario di Monte Sant'Angelo, Via Cintia, 80126 Napoli, Italy}

\author{Fabio Sciarrino }

\email{fabio.sciarrino@uniroma1.it}

\address{Dipartimento di Fisica, Sapienza Universit\`{a} di Roma, Piazzale Aldo Moro 5, I-00185 Roma, Italy}

\begin{abstract}
Engineering single-photon states endowed with Orbital Angular Momentum (OAM) is a powerful tool for quantum information photonic implementations. Indeed, thanks to its unbounded nature, OAM is suitable to encode qudits allowing a single carrier to transport a large amount of information. Nowadays, most of the experimental platforms use nonlinear crystals to generate single photons through Spontaneous Parametric Down Conversion processes, even if this kind of approach is intrinsically probabilistic leading to scalability issues for increasing number of qudits. Semiconductors Quantum Dots (QDs) have been used to get over these limitations being able to produce on demand pure and indistinguishable single-photon states, although only recently they were exploited to create OAM modes. Our work employs a bright QD single-photon source to generate a complete set of quantum states for information processing with OAM endowed photons.
We first study the hybrid \textit{intra}-particle entanglement between the OAM and the polarization degree of freedom of a single-photon. We certify the preparation of such a type of qudit states by means of the Hong-Ou-Mandel effect visibility which furnishes the pairwise overlap between consecutive OAM-encoded photons.
Then, we investigate the hybrid \textit{inter}-particle entanglement, by exploiting a probabilistic two qudit OAM-based entangling gate. The performances of our entanglement generation approach are assessed performing high dimensional quantum state tomography and violating Bell inequalities. Our results pave the way toward the use of deterministic sources (QDs) for the on demand generation of photonic quantum states in high dimensional Hilbert spaces.%, providing an efficient state of the art platform with application in various fields ranging from communication to study of quantum mechanics foundation.
%, such as photons states characterized by a maximum overlapping and entangled states. The capability of generating suitable states has been certified through Hong-Ou-Mandel (HOM) measurements between pairs of single-photon states encoded in superposition of OAM and polarization degrees of freedom. Then, we investigate both \textit{intra}-particle entanglement, between the OAM and the polarization degree of freedom of a single-particle, and \textit{inter}-particle one, by exploiting a probabilistic two qudit OAM-based entangled  gate. The performances of our entanglement generation approach are assessed performing high dimensional quantum state tomography and violating Bell inequalities. Our results pave the way toward the application of deterministic sources (QDs) for the on demand production of photonic quantum states in high dimensional Hilbert spaces.%, providing an efficient state of the art platform with application in various fields ranging from communication to study of quantum mechanics foundation.
\end{abstract}

\maketitle

\section{Introduction}
\label{sec:1}
In the last decades, structured light states characterized by an on-demand distribution for both field amplitude and phase have gained great interest \cite{rubinszteindunlop2016roadmap}. Among them, twisted beams carrying Orbital Angular Momentum (OAM) have been the focus of several studies due to their wide range of applications. As pointed out by Allen \textit{et al.} \cite{allen_0AM_1992}, OAM is carried by all the beams that present a phase term of the form $e^{i \ell \phi}$ where $\phi$ is the azimuthal angle in cylindrical coordinates and $\ell$ an unbounded integer. This phase term is responsible for the typical helicoidal wavefront, and each photon shows an OAM equal to $\ell \hbar$.

In classical domain, the non trivial phase structure of OAM states is used in several protocols covering a wide number of fields such as metrology \cite{lavery2013detection}, imaging \cite{Torner:05, Simon2012, Uribe-Patarroyo2013}, particle trapping \cite{Zhan} and communication \cite{willner2015optical, malik2012influence, wang2012terabit, baghdady2016multi, bozinovic2013terabitscale, gibson2004free,krenn_2016}. The unbounded nature of the OAM is instead the basis of its employment in quantum information. Therefore, OAM modes are used in quantum communication \cite{Cozzolino_rev,krenn2015twisted, Malik2016, cozzolino2019air, zhou2019using, Wang2015}, cryptography \cite{Mirhosseini_2015, Sit17, Bouchard_18}, simulation \cite{cardano2016statistical, cardano_zak_2017, Buluta2009}, computation and information processing \cite{Lanyon2009, ralph2007efficient, hiesmayr2016observation}. In particular, OAM-based encoding enlarges the amount of information that a single-photon can support, leading to increased security in the communication protocols \cite{bechmannpasquinucci2000quantum,sheridan2010security}. 
When the helicoidal wavefront is coupled with a nontrivial distribution of the Spin Angular Momentum (SAM), also known as polarization, a new class of states called Vector Vortex (VV) is introduced. %Given the peculiar coupling between the two degrees of freedom, these states turn out to be intrasystem maximally entangled in the OAM and polarization degrees of freedom.
Given this peculiar coupling, VV beams turn out to be \textit{intra}-system maximally entangled in the OAM and polarization degrees of freedom.
%Coupling the OAM with the other component characterized the total angular momentum of single photon, the Spin Angular Momentum (SAM) also known as polarization, a new class of states called Vector Vortex Beams (VVBs) gives rise. These are characterized by a non trivial polarization distribution along the beam profile. 
As for the OAM modes, VV beams are applied in several areas both in classical and quantum regime such as optical trapping \cite{li2017transverse, Cardano2015Rev}, communication \cite{vallone_qkd_2014, dambrosio2012complete}, computing \cite{Slussarenko2009,Nagali:09, Nagali2009,Deng:07,oamchip,Silva_2016,parigi2015storage, giordani2020machine,Pinheiro:13}, sensing and metrology \cite{kozawa2018superresolution, suprano2020propagation,polino2020photonic,fickler2012quantum, berg2015classically}.
Moreover, knowing the importance of the Hong-Ou-Mandel (HOM) effect \cite{HOM} and its applicability in quantum information science \cite{Bouchard_2020}, the interference behaviour between structured photons has also been studied in order to perform increasingly complex tasks \cite{Karimi2014Hom,Zhang2016,Hiekkam2021}.

Despite the large number of applications, sources that  produce single photons carrying OAM deterministically and with high brightness are still under development \cite{chen2021bright}. In fact, most of the experimental implementations leverage on producing single photons through Spontaneous Parametric Down Conversion (SPDC) in nonlinear crystals and modulating their states using bulk systems such as Spatial Light modulators (SLMs) \cite{fickler2013real, Yao:06} and q-plates \cite{marrucci-2006spin-to-orbital,Marrucci2011Rev,sit2017high}. However, SPDC is intrinsically probabilistic and suffers from a trade-off between the brightness and the purity of the produced single photons. Moreover, since in each process it is always possible to generate more than one photon, these kinds of sources undermine the security of quantum cryptography schemes \cite{brassard2000limitations}. 
%To overcome these limitations, quantum dot (QD) or nitrogen-vacancy based quantum emitters have been proposed and implemented \cite{chen2021bright,schanne2020spontaneous,wu2022room}. %QUI METTEREI IN PARTICOLARE I QD COSA SONO E CHE LA MAGGIORPARTE DEI LAVORI SI CONCENTRANO SULLA PREPARAZIONE DI STATI A SINGOLI FOTONI (VEDI DA MAURO). RECENTEMENTE ALCUNI HANNO CERCATO DI GENERARE SINGOLI FOTONI CON OAM (TUE CITAZIONI).  
%Articles I found that use quantum dots and generate OAM states and vector vortex beams \cite{chen2021bright,schanne2020spontaneous}
Semiconductor Quantum Dots (QDs) have emerged as a platform to overcome these limitations. Acting as artificial atoms when resonantly pumped with pulsed lasers, QDs are capable of generating indistinguishable single photons with high brightness in a nearly-deterministic fashion \cite{somaschi2016, wang2019, uppu2020, tomm2021}. However, most of the effort was concentrated on the generation of single or entangled states encoding the information in the photons polarization \cite{waks2002quantum,collins2010quantum, Li_2020,Istrati2020,Jin-Peng2021,dousse2010ultrabright} or in the temporal domain \cite{takemoto2015quantum}. Only recently, works exploiting QDs to engineer OAM modes \cite{chen2021bright} %and VVBs \cite{schanne2020spontaneous} 
within a prepare-and-measure framework have appeared. In particular, integrated sources based on microring resonators embedded with QDs \cite{chen2021bright} have been implemented, in which the OAM states encoded in the generated single photons could not be easily manipulated.
%and Au spiral structures covered by layers of colloidal QDs \cite{schanne2020spontaneous} were implemented.

%and the produced state is analyzed by imaging approaches or through projective measurement via a spatial light modulator.\\

\begin{figure}[ht!]
    \centering
    \includegraphics[width=0.5\textwidth]{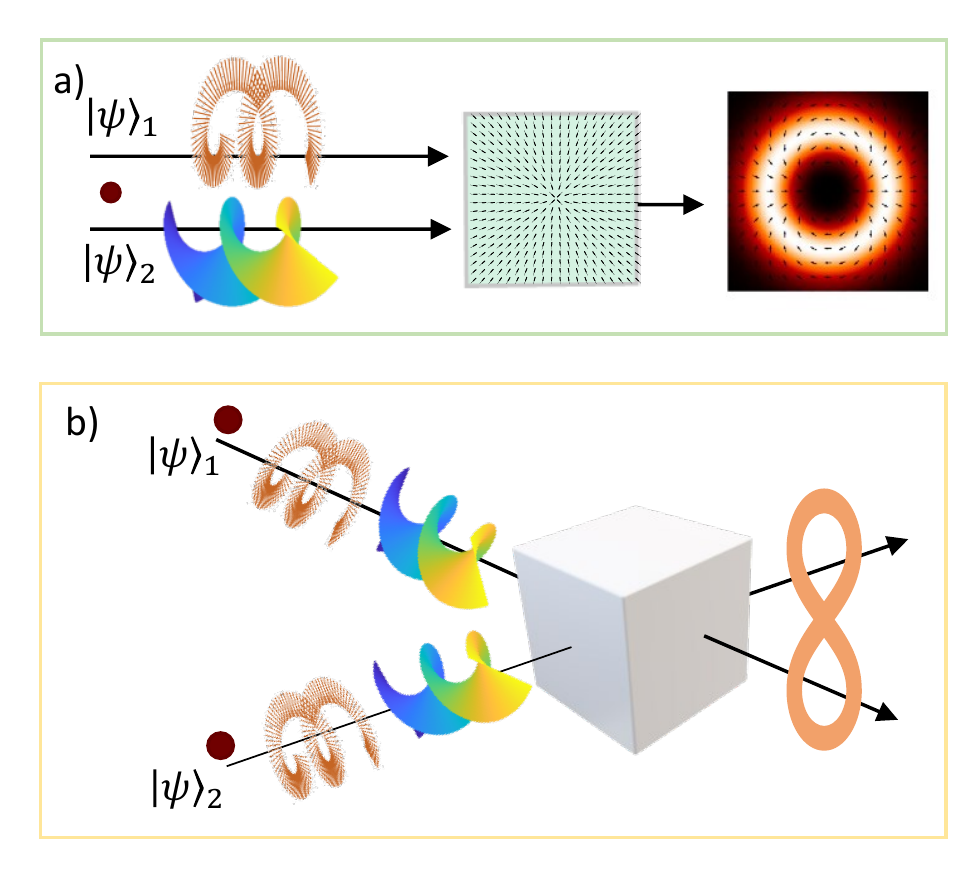}
    \caption{\textbf{Entanglement generation.} %a) Circuital representation of the entangling quantum gate. A Hadamard port is employed to initialize the first qubit in the balanced superposed state, while the controlled unitary gate correlates the two subsystems creating a maximally entangled bipartite quantum state. 
    a) In the \textit{intra}-particle entanglement, the polarization and OAM subsystems are made to interact using a q-plate. The two-dimensional state $\ket{\psi}_1$ is initialized with the right polarization $\ket{R}=\ket{0}$, while the qudit $\ket{\psi}_2$ is prepared with a null OAM value $\ket{0}$. The action of the unitary operator consists of increasing or decreasing the OAM value in a polarization-dependent way. b) In the \textit{inter}-particle regime, two photons characterized by defined states in the hybrid space composed of polarization and OAM interfere using a beam-splitter. Fixing the elements of the computational basis as $\ket{0}=\ket{L,-2}$ and $\ket{1}=\ket{R,2}$, both $\ket{\psi}_1$ and $\ket{\psi}_2$ are initialized with the qubit state $\ket{0}$, and after post selecting on the coincidence counts a probabilistic entangling quantum gate is implemented. It is worth noting that considering separately the polarization and OAM Hilbert spaces of both photons, the proposed apparatus implements a 4-qubits gate. %(see Supplementary Information).
    }
    \label{fig:sch_fig}
\end{figure}

%At variance with \cite{chen2021bright}, instead of designing compact sources and assessing their performances, here we focus on the development of quantum information processing protocols.
At variance with \cite{chen2021bright}, we exploit commercial QD based single-photon sources, and focus on the development of quantum information processing protocols with VV beams.
%In this work, we focus on the generation of indistinguishable single photons endowed with the OAM degree of freedom. By appropriately manipulating the polarization of single photons generated by a QD source and employing inhomogenous and birefringent media, known as \textit{q-plate} \cite{marrucci-2006spin-to-orbital,Marrucci2011Rev}, we are able to encode different OAM states. The indistinguishability of such states have been verified by sending single photons at each input of a 50:50 beam splitter and, thus, evaluating an Hong-Ou-Mandel (HOM) interference visibility. Moreover, the same apparatus has been exploited to implement a probabilistic 2-qudits quantum gate able to generate entangled photon pairs with dimension up to 4.
Specifically, well-known OAM manipulation technologies have been extensively used to develop high-dimensional quantum communication protocols \cite{cozzolino2019air, Sit17}, to reach a high flexibility in engineering arbitrary qudit states \cite{giordani_2018, suprano2021dynamical}, and to develop simulated processes based on the quantum walk dynamics \cite{cardano2016statistical, cardano_zak_2017}. Here, we combine these technologies with an innovative and nearly deterministic single-photon source, opening the way for further developments of quantum information protocols that take advantage of high-dimensional resources and of the benefits introduced by using QDs. In particular, besides focusing on interfacing between these two kinds of technologies, we perform a step forward and study the hybrid entanglement in high-dimensional Hilbert spaces implementing a quantum gate both in the \textit{intra-} and \textit{inter-} particle regime (Fig. \ref{fig:sch_fig}). Previously, states characterized by hybrid  \textit{intra}-photon entanglement between the OAM and polarization degree of freedom have been generated via SPDC processes \cite{cozzolino2019air,Nagali:09,D'ambrosio2016,Karimi2010}. However, since this kind of source is probabilistic, the state is certified in an heralded configuration which drastically decreases the generation rate. On the contrary, the employment of a deterministic single-photon source, allows us to certify the state directly on the single counts increasing the generation rate. Moving toward the \textit{inter}-particle regime, versatility and flexibility in the generation and manipulation of indistinguishable photons are crucial features for gate implementation. We then move a step forward with respect to Ref. \cite{chen2021bright} by investigating the indistinguishability of the generated photons and employing a versatile approach. In our platform, the combination of QD based single-photon sources and well-known OAM manipulation devices allow us to satisfy the aforementioned requirements. 

This work is organized as follows. We start by studying the single-photon \textit{intra}-particle entanglement generation in VV states. %through by appropriately manipulating its polarization and exploiting a %suitable device called 
By means of
q-plate devices \cite{marrucci-2006spin-to-orbital,Marrucci2011Rev}, we couple the two components of the angular momentum degree of freedom and generate VV beams %states 
(Fig. \ref{fig:sch_fig}-a). Then, we move to the multi-photon scenario. Preliminary, we certify the efficiency of encoding OAM states on single photons emitted by the QD in different pulses of the pump beam % and manipulated using waveplates and q-plates, 
through the evaluation of Hong-Ou-Mandel interference visibility. Thus, we implement a 2-photons probabilistic quantum gate able to generate OAM-based entangled photon pairs involving up to 4 subsystems (Fig. \ref{fig:sch_fig}-b). We verify both the single-photon \textit{intra}-particle and the two-photon entanglement performing quantum state tomography and evaluating the Bell inequality in the Clauser–Horne–Shimony–Holt (CHSH) fashion. %encode different OAM states. %We assess the indistinguishability of such states sending single photons at each input of a 50:50 beam splitter and, thus, evaluating an Hong-Ou-Mandel (HOM) interference visibility. After that, we exploit the same apparatus to implement a probabilistic 2-qubit quantum gate capable of generate entangled photon pairs with dimension up to 4. 

\begin{figure}[!ht]
    \centering
    \includegraphics[width=0.5\textwidth]{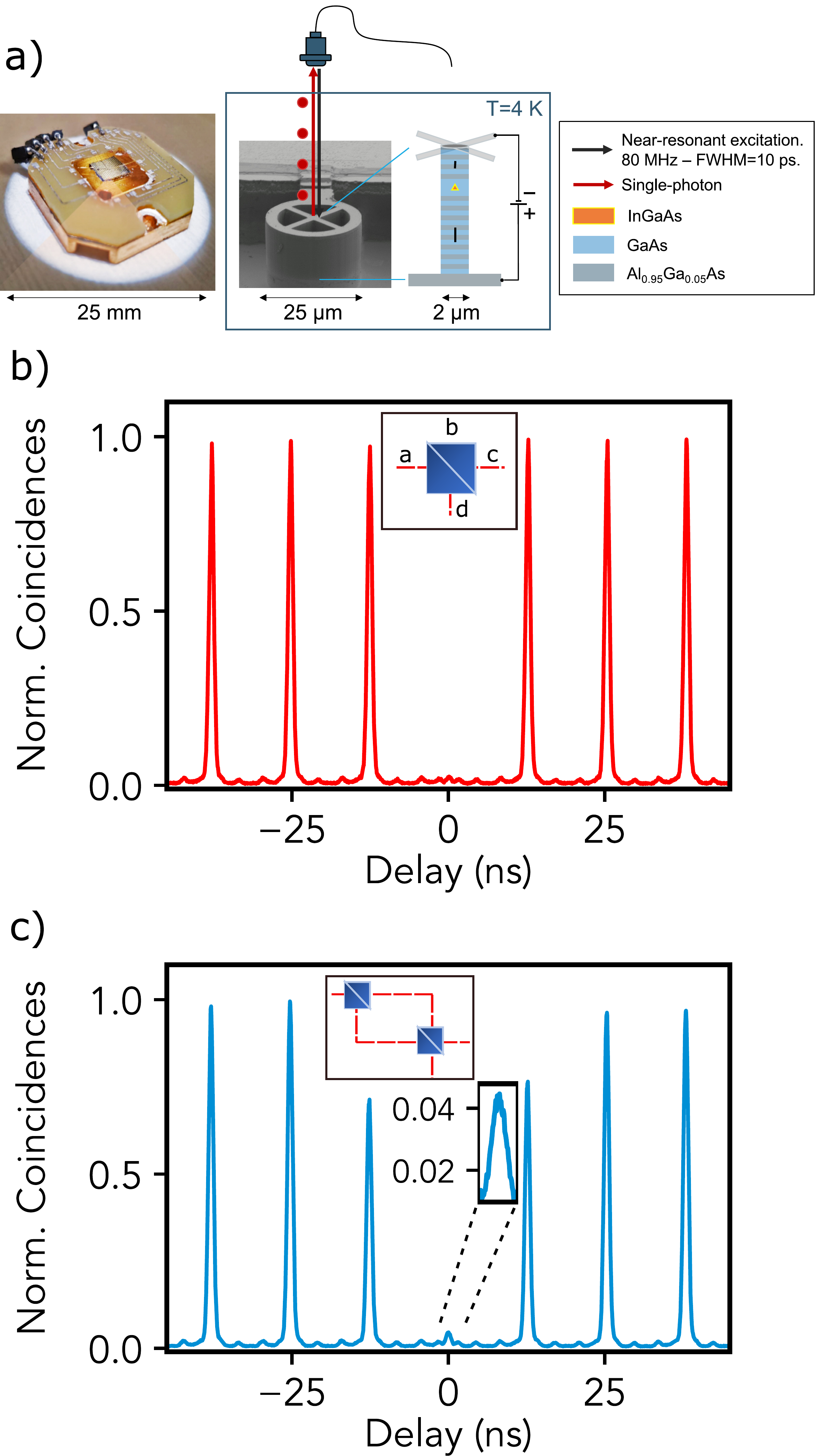}
    \caption{\textbf{Source Hong-Ou-Mandel interference and second-order correlation function.} a) The single-photon source (left) is a commercial device (Quandela): InGaAs quantum-dot based bright emitters are embedded in electrically-contacted micropillars (right). The source is pumped with a near-resonant ($\Delta\lambda$=-0.6~nm) FWHM 10~ps 79~MHz-pulsed laser (red arrow). The single photons (red dots) are emitted at a wavelength of 927.8~nm and are directly coupled to a SMF.
    b) Through a standard Hanbury Brown and Twiss setup, we measure the second-order autocorrelation histogram of our QD-based source as a function of the delay. We obtain a single-photon purity of $g^{(2)}(0) =( 1.26 \pm 0.05) \%$. 
    c) Normalized correlation histogram, obtained via a HOM interference experiment, through which we measure an 2-photon interference fringe visibility between subsequent single photons emitted by the QD source of $V_{HOM} = (93.05 \pm 0.06)\%$. Moreover, following Ref. \cite{ollivier2021hong}, we obtain an indistinguishability value of $M_s = (95.5 \pm 0.1) \%$. }
    \label{fig:my_label}
\end{figure}

\section{Experimental platform}
\label{sec:exp_setup}
In this section, we preliminarily describe the employed quasi-deterministic single-photon source, evaluating the intensity auto-correlation and indistinguishability of the generated photons. Subsequently, we present the implemented scalable platform in which, by interfacing well-known OAM manipulation devices with the QD source, entangled \textit{intra}- and \textit{inter}-particle states are generated in the hybrid Hilbert space composed of OAM and polarization.

\subsection{Single-photon source}\label{sec:source}

The single-photon source is a quantum dot (QD) based emitter embedded in an electrically controlled cavity on a commercially available \textit{Quandela} \textit{e-Delight-LA} photonic chip. A single self-assembled InGaAs QD is surrounded by a two Braggs reflectors made of GaAs/Al$_{0.95}$Ga$_{0.05}$As $\lambda/4$ layers with 36 (16) pairs for the bottom (top) and positioned in the center of a micropillar \cite{somaschi2016}. The micropillar is connected to a larger circular structure that is electrically contacted enabling the tuning of the emission frequency of the QD with Stark effect. The sample is kept at 4~K in a low-vibration closed-cycle He cryostat \textit{Attocube - Attodry800}. 
%The QD source is pumped with a $79$~MHz-pulsed laser, whose spectrum is centered at $920$~nm with a 20~nm-large bandwidth; the laser pulses then pass through a \MP{Q-Shaper (Quandela)} 4f pulse shaper to select a specific wavelength and achieve a bandwidth $<100$~pm.
The QD source is pumped with a $79$~MHz-pulsed laser shaped with a QShaper (Quandela) 4f pulse shaper to select a specific wavelength and achieve a bandwidth of $\sim 100$~pm.
The optical excitation of the QD is achieved in an LA phonon assisted configuration with a laser at $927.2$~nm blue-detuned from the transition \cite{Thomas2021}, which enables single-photon generation by exciton emission at $(927.8\pm0.2)$nm (Fig. \ref{fig:my_label}-a). The emitted photons are directly coupled in single-mode fiber (SMF) %, then spectrally separated by the residual pumping laser with a sequence of three bandpass filters ($< 0.8$nm) in free-space and coupled again in a SMF.
and spectrally separated from the residual pumping laser with bandpass filters.
At the output of the \textit{e-Delight-LA} system, we measure a single-photon count rate of $R_{\mathbf{det}} = ~4$ MHz. The fibered brightness of the single-photon source depends mainly %by the photon generation probability per pump pulse, the extraction efficiency from the cavity, 
on the coupling efficiency into the SMF, the spectral separation transmission of the single-photon stream from the pump laser - whose effects we estimate in an overall efficiency of $\eta_{\mathbf{setup}} \sim 52\%$ - and the detector efficiency, estimated to be around $\eta_{\mathbf{det}} \sim 38\%$. Using this figures, we estimate a \textit{first lens brightness} of $B = \frac{R_{\mathbf{det}}}{R_{\mathbf{exc}}\eta_{\mathbf{det}}\eta_{\mathbf{setup}}} \sim 26\%$, where $R_{\mathbf{exc}}$ is the pump frequency. %\commDZ{Definire $R_{exe}$}
The overall quality of the single-photon generation can be characterized by measuring the multi-photon emission and indistinguishability. Using a standard Hanbury Brown and Twiss setup, we measured an second-order auto-correlation of $g^{(2)}(0) =( 1.26 \pm 0.05) \%$. Such figure is computed by normalizing the zero-time delay coincidences to the side peaks coincidences between two consecutive near-resonant excitations (Fig. \ref{fig:my_label}-b). We also measured the indistinguishability between photons successively emitted %$\Delta_{\tau}=12.3ns$
by the QD, through a Hong-Ou-Mandel (HOM) interference experiment \cite{hong1987measurement}. Two consecutively emitted photons are split by a beam splitter (BS) and coupled in SMFs, whose length is chosen to delay one of them by $\approx$12.5 ns to ensure temporal overlap on a second BS. At its outputs, photons are collected in Avalanche Photodiode Detectors (APDs) to record photons coincidence counts. Therefore, we evaluate a 2-photon interference visibility derived from the correlation histogram (Fig. \ref{fig:my_label}-c) as $V_{HOM} =  1 - 2\frac{C_{0}}{\langle C \rangle_{t\rightarrow \infty}}$, where $C_0$ are the counts when the two photons are synchronized and $\langle C \rangle_{t\rightarrow \infty}$ are the average peak counts for relative temporal delays larger than one repetition rate of the laser. We measure an interference visibility $V_{HOM} = (93.05 \pm 0.06)\%$, which can be corrected to account for unwanted multi-photon components \cite{ollivier2021hong}, resulting in a photon indistinguishability equal to $M_s = (95.5 \pm 0.1) \%$.  %$V^*_{HOM} = (95.52 \pm 0.096) \%$.

\subsection{Experimental implementation of OAM-based platform}
\begin{figure*}[ht!]
    \centering
    %\includesvg[width=1\textwidth]{Setup_exp_71}
    \includegraphics[width=1\textwidth]{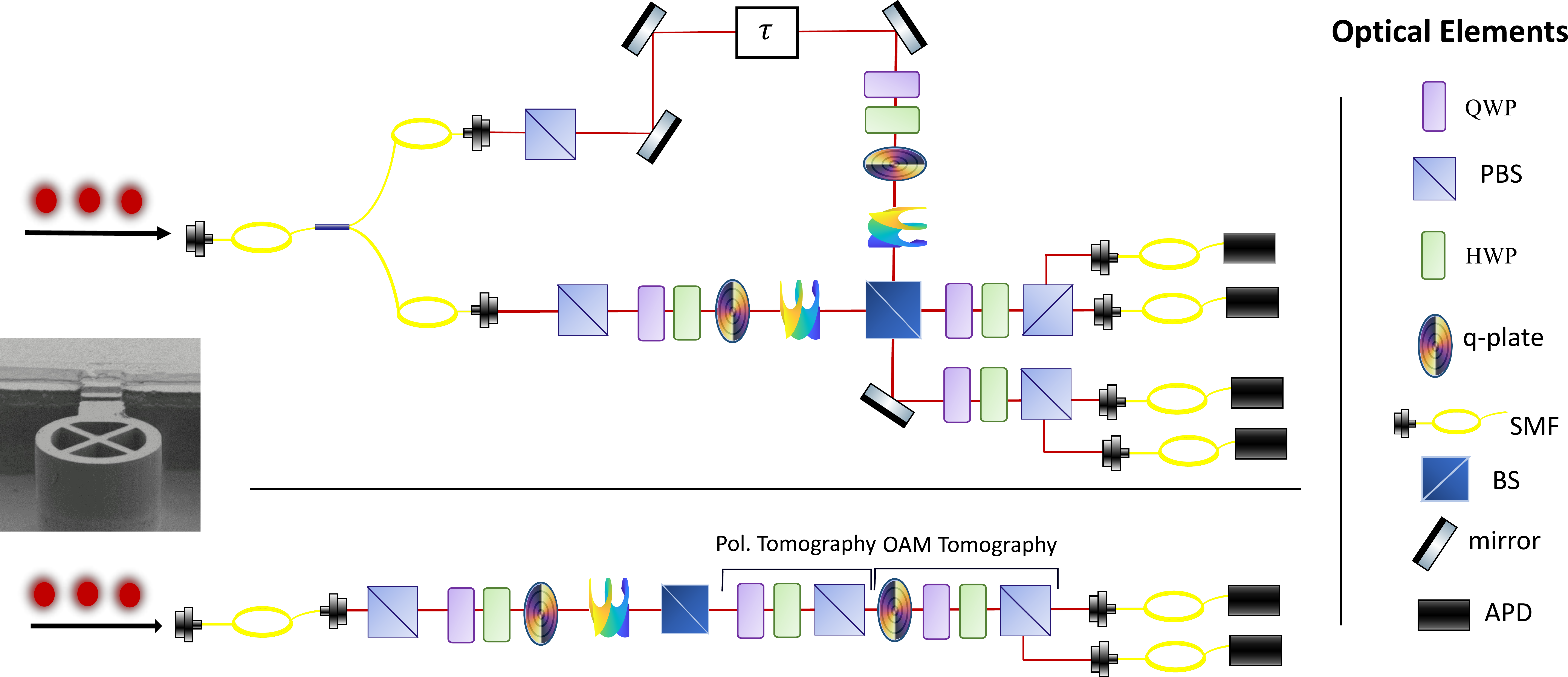}
    \caption{\textbf{Experimental Setup.} Single-photon states at a wavelength of $927.8 \pm 0.2$ nm are generated using a QD source pumped with a shaped 79 MHz-pulsed laser at $927.2$ nm. Then a fiber-BS splits the photons between the two arms of the setup, and after passing through a PBS the input states have horizontal polarization and OAM eigenvalue $m=0$.
   In both paths, series of QWP, HWP and q-plate are used to produce OAM modes of the form reported in Eq. \eqref{eq:VVB}, while in one of the arms, a delay line ($\tau$) is inserted in order to synchronize on the BS the photons emitted in different pulses of the pump beam. The \textit{intra}-particle regime is investigated removing the fiber-BS and performing all the experiment on a single line, involving the first input and output of the BS, as shown in the below panel.
   On the other hand, in the \textit{inter}-particle experiment, the photons are sent to the fiber-BS and the gate is implemented interfering on the second BS. After passing through the BS the state of the photons is analyzed, coupled to SMFs and detected by APDs. The measurement setup consists in two different stages, a series of q-plate, QWP, HWP and PBS is used to study the OAM states of the photons coupled with the polarization, while a QWP, HWP and a PBS compose the polarization analysis setup. In the \textit{inter}-particle regime only OAM analysis is performed on the photon pairs. While, in the \textit{intra}-particle regime both analysis setups are used to separately investigate the polarization and OAM content of single photons, as shown in the below panel.}
    \label{fig:Exp_setup}
\end{figure*}

We experimentally implemented a flexible platform for the study of single and multi-photons properties% such as entanglement and indistinguishability
, and capable of implementing a probabilistic entangling quantum gate.
%capable of both measuring the visibility of HOM interference between quantum OAM states and implementing a 2-qubit probabilistic quantum gate
 A visual scheme of the setup is reported in Fig. \ref{fig:Exp_setup}.
 
 For this purpose, the stream of single photons generated by the QD is preliminarily split through a fiber-BS and OAM encoding is performed separately on the two outputs. In particular, the input state, $\ket{H,0}$, having horizontal polarization and null OAM value, is selected through single mode fibers and polarizing beam splitters (PBS). In the engineering stage, by placing a set of waveplates together with a q-plate on each arm, we are able to independently generate two distinct OAM-encoded single-photon states. In particular, a q-plate is a thin film of birefringent material (in our case, nematic liquid crystals) characterized by a non-uniform distribution of the optic axis across the plane transverse to the light propagation direction. The angle between the optic axis and the horizontal axis $x$ of the device follows the relation $\alpha(\phi)=\alpha_0+q \phi$, where $\phi$ is the azimuthal angle in the transverse plane, $\alpha_0$ is the optic axis orientation for $\phi=0$ and $q$ is the topological charge, i.e.\ the winding number of the optic axis for $\phi \in [0,2\pi]$. Owing to the inhomogeneity of its optic axis distribution and to the resulting Pancharatnam-Berry geometric phases, the q-plate develops on the OAM degree of freedom of single photons an action that depends on their polarization, according to the following expression \cite{marrucci-2006spin-to-orbital,Marrucci2011Rev}:
\begin{equation}
    \hat{Q}=\sum_m |m-2q\rangle \langle m|\otimes |L\rangle \langle R|+ |m+2q\rangle \langle m|\otimes |R\rangle \langle L|,
    \label{eq:qp_action}
\end{equation}
where $\ket{R}$,$\ket{L}$ indicate respectively right and left circular polarization states and $\ket{m}$ represents the OAM value. 

Therefore, an optical setup consisting of a quarter-waveplate (QWP), a half-waveplate (HWP) and a q-plate with $q=1$, acting on the input state $\ket{H,0}$, is able to engineer arbitrary superpositions of $\ket{L,-2}$ and $\ket{R,2}$ as given by:
\begin{equation}
    \ket{\Phi}=\cos(\theta/2)\ket{L,-2}+e^{i \psi} \sin{(\theta/2)}\ket{R,2}
    \label{eq:VVB}
\end{equation}
where, $\theta \in [0,\pi]$ and $\psi \in [0,2\pi]$ can be set by properly orienting the optic axes of QWP and HWP. In this way, \textit{intra}-system entanglement between OAM and SAM degrees of freedom of single photons can be easily achieved. In particular, for $\theta=\pi/2$, the superpositions given in Eq.\ (\ref{eq:VVB}) correspond to the above mentioned VV states.

\begin{figure*}[ht!]
    \centering
    \includegraphics[width=\textwidth]{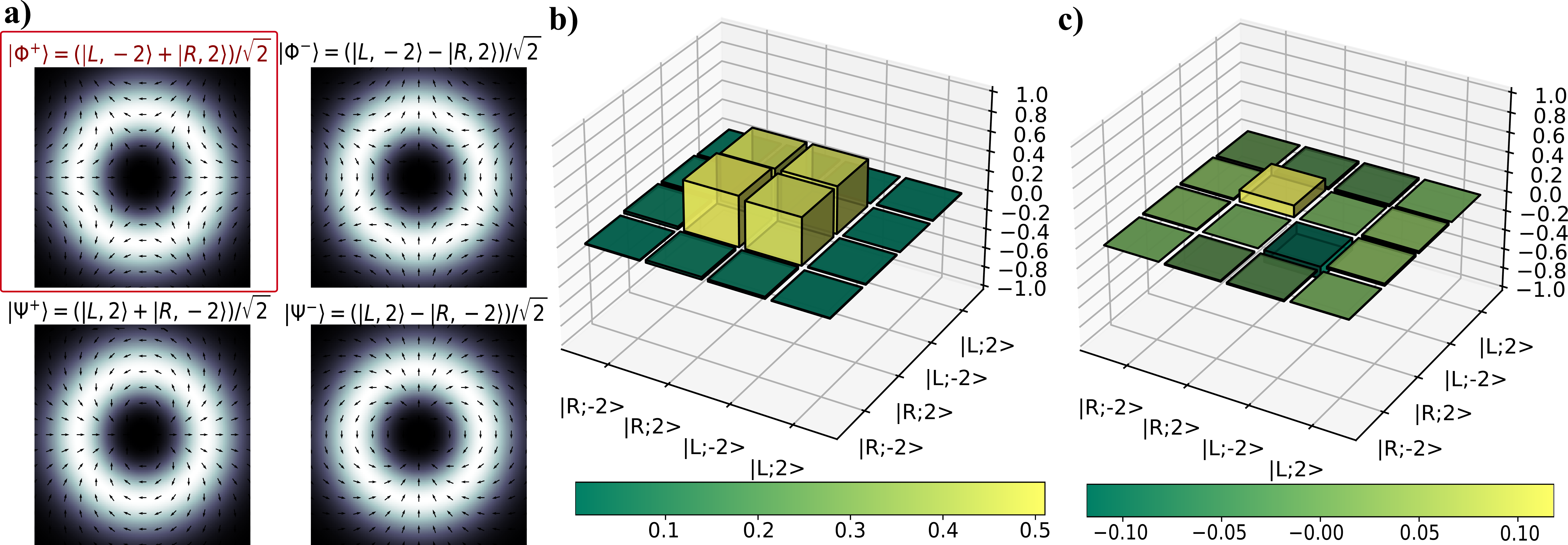}
    \caption{\textbf{\textit{Intra}-particle entangled state:} (a) Intensity and polarization patterns of the Bell states basis in the combined OAM and polarization space. As highlighted by the red box, we focused our attention on the $\ket{\Phi^{+}}$ state. (b) Real and (c) imaginary  parts of the measured density matrix for the $\ket{\Phi^{+}}$ state reconstructed via quantum state tomography. The fidelity between the reconstructed state and the theoretical one is equal to $\mathcal{F}=0.9714 \pm 0.0007$, where the standard deviations are estimated through a Monte Carlo approach assuming a Poissonian statistics. }
    \label{fig:intraparticle}
\end{figure*}

Subsequently, the two arms are synchronized by introducing a fixed delay in fiber and a tunable delay in air and then sent to a bulk beam splitter (BS) used to probabilistically generate an entangled state between the two photons in the hybrid space of OAM and SAM by a postselection on the measured events.

Finally, in both \textit{intra}- and \textit{inter}- experiment the state reconstruction is performed by using q-plates and polarization tomography setups comprising a QWP and a HWP followed by a PBS. In fact, the OAM tomography setup is implemented by adding a q-plate in front of the polarization tomography setup to convert the correlations present in the OAM degree of freedom on the polarization space, as can be evinced from Eq. \eqref{eq:qp_action}. 
In particular, in the \textit{intra}-particle regime, the fiber-BS is removed and the entangled state is generated along the lower arm of the interferometer. Both the polarization and OAM analysis of such states is performed along one BS output, by inserting the polarization tomography setups followed by the OAM tomography one. Instead, the analysis of the \textit{inter}-particle entangled state is performed by placing only the OAM tomography setups on each BS output.
After the projection, the photons are collected in SMFs and detected using APDs.
%so the same three elements are exploited after that.  After the projection, the photon are collected in Single Mode Fibers (SMFs) and detected using Avalance Photodiode Detectors (APDs).
This scheme is used both to study the entanglement content of the states through Bell inequalities violation and to perform quantum state tomographies.
%Specifically, to collect photons in single-mode fibers (SMFs), a q-plate with $q=1$ is employed to obtain quantum states with $m=0$, while polarization tomography is performed through a quarter-and-a-half wave plate followed by a PBS. \add{After the coupling in SMF, the photons are measured through Avalance Photodiode Detectors (APD) placed at the two outputs of the setups.} \add{This measurement stage can be used to retrieve the quantum states description in the combined space of OAM ad polarization, allowing us to verify their entanglement content.}

%Losses setup:
%\begin{itemize}
 %   \item 2 Mating Sleeve: $0.89^2$
 %  \item 2 q-plates: $0.7^2$
 %   \item loss for BS-interference: $0.5$
 %   \item Initial PBS loss for unpolarized emission??
%\end{itemize}
%$\eta_{setup}=0.89^2\times0.7^2\times 0.5=0.194$.
%\\
%\\
%Losses source: $4/80=0.05$
%\\
%\\
%Losses Tot: $0.05\times 0.194=0.0097$.
%\\
%\\
%Coincidences: $\frac{80 \times (0.0097)^2}{2}$ MHz=$3.76$ KHz. 

\section{Entanglement certification}
%\commDZ{Questo è solo un titolo provvisorio per evidenziare cosa studiamo con il setup, non mi piace troppo se vi viene in mente qualcosa di meglio cambiatelo. Prima era "Theoretical Background" ma per me qui si possono riportare anche i risultati, non farei un'ulteriore sezione}

In this section, we provide the theoretical description and report the results obtained studying the \textit{intra}-particle and \textit{inter}-particle hybrid entanglement generated with the experimentally implemented platform. In all cases of interest, entanglement is certified through a violation of a CHSH Bell inequality and complete state tomography.

\subsection{Vector vortex beam: \textit{intra}-system entanglement}
%For studying the intraparticle entanglement we focus our attention only on one of the arms of the setup. 
The first investigation regards the generation of VV beams encoded into the single-photon states generated by the QD source. The VV beams are superpositions of two or more different OAM beams associated to orthogonal circular polarizations, an example is given in Eq. \eqref{eq:VVB}. Here, the two systems individuated by the OAM eigenstates $\{\ket{-2},\ket{2}\}$ and the polarization states $\{\ket{R},\ket{L}\}$ can be exploited for encoding two qubits. In this way, it is possible to define a complete basis of maximally entangled states between these two degrees of freedom. The set of Bell-like states is reported in Fig.\ref{fig:intraparticle}, in which the non-uniform polarization distribution in the transverse plane is highlighted.

%\textcolor{red}{In our setup, we select the signal form one of the two arms of the first BS after the QD source (see Fig.\ref{fig:Exp_setup}).}
In our setup, to increase the generation rate, the signal is sent only in one of the two arms of the interferometer by removing the first fiber-BS (see Fig.\ref{fig:Exp_setup}). The VV beams are prepared by making horizontally polarized photons passing subsequently through a QWP, a HWP and a q-plate with $q=1$. In this way, the state produced by the device is described by $\theta = \pi/2$ in Eq. \eqref{eq:VVB} and a value of $\psi$ which depends on the $\alpha_0$ of the q-plate optic axis. This additional phase term is compensated by a further HWP (not shown in Fig. \ref{fig:Exp_setup}) %. Correcting the phase factor with the incoming waveplates 
in order to have $\psi = 0$. The final entangled state between OAM and polarization will be

\begin{equation}
    \ket{\Phi^{+}}= \frac{1}{\sqrt{2}}\left[\ket{L,-2}+ \ket{R,2}\right]
    \label{eq:VVB_intra}
\end{equation}

%Relabeling both polarization $\{\ket{L},\ket{R}\}$ and OAM basis $\{\ket{2},\ket{-2}\}$ as the qubit computational basis $\{\ket{0},\ket{1}\}$, 
%The state in Eq. \ref{eq:VVB_intra} is equivalent to a triplet Bell state as expected. %Therefore, this state is a maximally entangled state in two distinct degrees of freedom of single photons, respectively the OAM and SAM. 
Although such entanglement structure is not associated with non-local properties since it is encoded in a single carrier, these correlations can be detected using Bell-like inequalities. We refer to such type of quantum correlations as \textit{intra}-particle entanglement. 

%Moreover, by employing 
The adoption of a nearly-deterministic single-photon source allows us to perform the \textit{intra}-particle analysis without the need of heralding measurements or post-selection. The latter are unavoidable procedures for generating single-photon states with high purity via probabilistic sources.   %detecting the singles counts without the need for coincident measurements. 
This reduces drastically the losses allowing to reach a rate of $\simeq 99$ kHz of VV states generation (see Supplemental Information for further details). %\commTG{Domada: visto che c'è un BS all'inizio e uno dopo la generazione, il rate effettivo dovrebbe essere 4 volte quello riportato? Oppure 55khz già tiene conto dei BS?} \textcolor{blue}{Alessia: dovrebbe essere due volte perchè quello in fibra l'ho rimosso. Dici che metto quello?} \commDZ{Ho modificato la parte iniziale per dire che non c'è il fiber BS. Se vi va bene io lascerei così e metterei il valore doppio.}. %In particular, the correlations between the two degrees of freedom shown in Eq. \ref{eq:VVB_intra} has been investigated exploiting the measurement stage described in Section \ref{sec:exp_setup}. 
The quality of the state and of its entanglement structure has been certified  by the measurement stage setup shown and described in Section \ref{sec:exp_setup}. In particular,  we performed a quantum state tomography by analysing the OAM and the polarization independently via cascaded measurement stages as in the below panel of Fig. \ref{fig:Exp_setup}. The resulting density matrix is reported in Fig. \ref{fig:intraparticle} and the relative fidelity, computed by subtracting for dark counts, is $\mathcal{F}=0.9714 \pm 0.0007$. Moreover, we also certified the \textit{intra}-particle entanglement %present the OAM and the polarization degrees of freedom 
by evaluating a CHSH-like inequality. Collecting data for $20$ s, we obtained a raw violation of $S^{(raw)}=2.736 \pm 0.008$ which exceeds the separable bound by $92$ standard deviations, while the value obtained by subtracting the background signal is $S=2.792 \pm 0.008$ which exceeds the classical bound by $99$ standard deviations. The results are summarized in table \ref{Tab:Fid}.
%\commDZ{Domanda: Secondo voi sarebbe utile mettere descrivere tipo nelle SI che tipo di misure abbiamo fatto qui e riportarne anche i valori? (Visto che ci sono delle SI forse questo darebbe più completezza al tutto)}

\begin{figure*}[ht!]
    \centering
    \includegraphics[width=\textwidth]{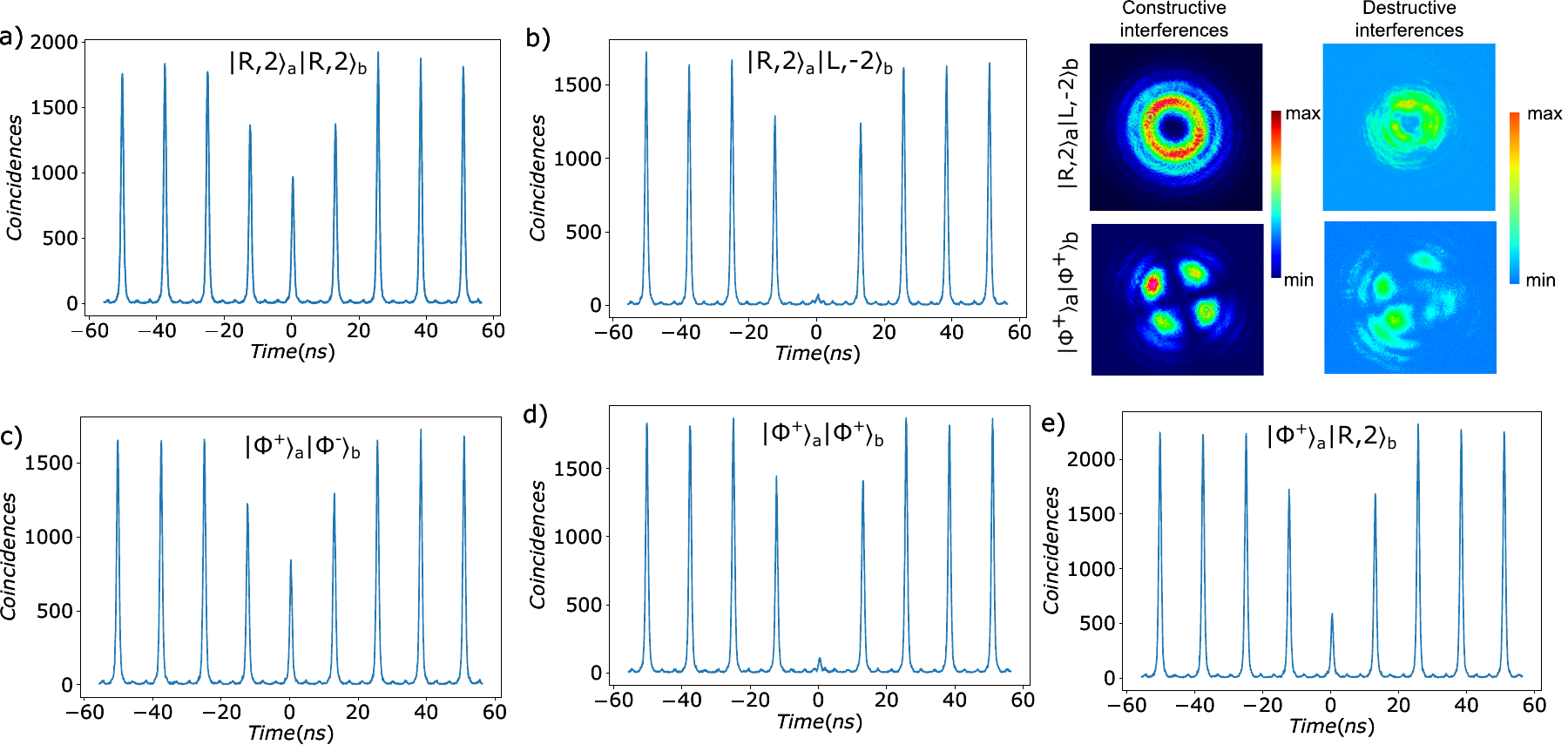}
    \caption{%\commTG{Se vi convince la nomenclarura degli stati nel testo e la versione senza formule andrebbe un po' rivista la caption e i nomi degli stati nelle figure.}
    \textbf{Hong-Ou-Mandel interference for OAM states:} Measured coincidences at the output of the final BS, see Fig. \ref{fig:Exp_setup}, for different input states in the hybrid space of OAM and polarization. A perfect HOM interference can be obtained only when the photon states are indistinguishable from the point of view of the observer.
     By knowing the BS action on circular polarization and OAM (see Supplementary Information), we observe a near-unitary visibility when the photons
    are prepared in the same eigenstate of the BS reflection operation, or when the initial states 
    have opposite circular polarization and OAM value. Moreover, we also analyze the hybrid configuration in which one photon is prepared in the state $\ket{R,2}$ and the other in the VV state $\ket{\Phi^+}$. In the latter case, the expected number of coincidences is half of the one obtained for distinguishable photons. %In the insets, the intensity patterns of the output modes of a q-plate for circularly (up) and linearly (bottom) polarized input light are reported.
    In the inset, the intensity patterns associated to constructive and destructive interference are reported for both initial states $\ket{R,2}_a \ket{L,-2}_b$ and $\ket{\Phi^+}_a \ket{\Phi^+}_b$.  }
    \label{fig:HOM}
\end{figure*}

\subsection{Certification of photon states generation}
%In order to certify the two photons state generated by the gate, we start analyzing the indistinguishability between

%Another computational resource beyond the entanglement for quantum information processing in photonic platforms is photons indistinguishability. 
In quantum information processes, an important computational resource relies on the capability of manipulating multiple photons and making them interact. Therefore, in this section, we assess the capacity of codifying specific OAM states on photons generated by subsequent pulsed pumping of the QD. This is performed by evaluating the visibility of HOM interference in a beam-splitter which is equivalent to a pairwise overlap estimation in a SWAP test \cite{Garcia_swap_hom}. There are some previous examples of HOM experiments with single-photon states carrying OAM \cite{Nagali2009, Ndagano19}, but our tests are among the first to be applied to vector beams generated by a deterministic single-photon source.

Let us first briefly review the effect of an unbiased BS on the field %operators. consider two optical input modes $a$ and $b$ of a 50:50 beam splitter, that carry 
annihilation and creation operators, $\hat{a},\hat{a}^{\dagger}$ and $\hat{b},\hat{b}^{\dagger}$.
The relation between input modes $\{a,b\}$ and output modes $\{c,d\}$ can be expressed as (Fig. \ref{fig:my_label}):
\begin{eqnarray}
    &\hat{a}^{\dagger}\longmapsto \frac{1}{\sqrt{2}}\left( \hat{c}^{\dagger}-\hat{d}^{\dagger}\right)\nonumber \\
    &\hat{b}^{\dagger} \longmapsto \frac{1}{\sqrt{2}}\left( \hat{c}^{\dagger}+\hat{d}^{\dagger}\right).
    \label{eq:BS}
\end{eqnarray}
By considering two photons at the two inputs of the beam splitter, the signature of the interference is a change in the probability to detect photons in different outputs (see Supplementary Information). In particular, two photons are indistinguishable if their states, associated to each degree of freedom, is the same from the point of view of the observer. To approach this condition in our setup, a delay line is used to synchronize the photons in the temporal domain. This is mandatory because the two single photons are emitted by the QD at different times. However, when the photons are characterized by OAM value different from zero and superposed polarizations, it is necessary to take into account the effect of reflections. Indeed, in a physical beam-splitter the semi-reflective mirror flips the elicity of both OAM and polarization. In other words, after one reflection we have $\{\ket{R},\ket{L}\}\rightarrow\{\ket{L},\ket{R}\}$ and $\ket{\pm2}\rightarrow\ket{\mp2}$, while horizontal and vertical polarizations are eigenstates of this operation with eigenvalues of opposite signs. Then, we have that the creation operators are changed as follows:
\begin{eqnarray}
    &\hat{a}^{\dagger}_{R}, \hat{b}^{\dagger}_{R} \longmapsto \frac{1}{\sqrt{2}}\left(\hat{c}^{\dagger}_{R}-\hat{d}^{\dagger}_{L}\right),\;\; \frac{1}{\sqrt{2}}\left(\hat{c}^{\dagger}_{L}+\hat{d}^{\dagger}_{R}\right)\nonumber \\
    &\hat{a}^{\dagger}_{L}, \hat{b}^{\dagger}_{L} \longmapsto \frac{1}{\sqrt{2}}\left(\hat{c}^{\dagger}_{L}-\hat{d}^{\dagger}_{R}\right),\;\; \frac{1}{\sqrt{2}}\left(\hat{c}^{\dagger}_{R}+\hat{d}^{\dagger}_{L}\right)\nonumber \\
    &\hat{a}^{\dagger}_{m}, \hat{b}^{\dagger}_{m} \longmapsto \frac{1}{\sqrt{2}}\left(\hat{c}^{\dagger}_{m}-\hat{d}^{\dagger}_{-m}\right),\;\; \frac{1}{\sqrt{2}}\left(\hat{c}^{\dagger}_{-m}+\hat{d}^{\dagger}_{m}\right).
    \label{eq:reflection}
\end{eqnarray}

Since the indistinguishability of photons generated by the source has been already checked in Section \ref{sec:source}, here we are interested in computing the overlap between VV states encoded in different photons. As for the previous analysis, the OAM and polarization degrees of freedom are controlled through a series of QWP, HWP and q-plate placed in each arm of the interferometer. This allows us to prepare the desired state for each photon.\\
%\textcolor{red}{To correctly describe the observed behaviors, it is necessary to take into account the effect of reflections on circular polarization and OAM. Indeed, in a physical beam-splitter the semi-reflective mirror flips the elicity of both OAM and polarization. In other words, after one reflection we have $\{\ket{R},\ket{L}\}\rightarrow\{\ket{L},\ket{R}\}$ and $\ket{\pm2}\rightarrow\ket{\mp2}$, while horizontal and vertical polarizations are eigenstates of this operation with eigenvalues with opposite signs (see Supplementary Information). }
Considering the BS action in Eq. \eqref{eq:reflection}, we expect no interference when the two photons are prepared as $\ket{R, 2}_a\ket{R,2}_b$, since the reflected beam and the transmitted one in the outputs $c$ and $d$ will display orthogonal states. Conversely, the HOM effect occurs when the initial state is $\ket{R, 2}_a\ket{L,-2}_b$. The correlation histograms, obtained via a HOM interference experiment, for both input states $\ket{R, 2}_a\ket{R,2}_b$ and $\ket{R, 2}_a\ket{L,-2}_b$ are reported in Fig. \ref{fig:HOM}-a,b. The visibility of such HOM experiments quantifies the variation, from the maximum to minimum overlapping between the wavefunctions, of the probability to detect photons in different outputs. The obtained visibilities are %$V_{\ket{R,2},\ket{R,2}}=4.20 \pm 0.96 \%$ 
$V_{\ket{R,2},\ket{R,2}}=-4 \pm 1 \%$ and %$V_{\ket{R,2},\ket{L,-2}}=90.05 \pm 0.30 \%$
$V_{\ket{R,2},\ket{L,-2}}=90.1 \pm 0.3 \%$, respectively.  %\commTG{La formula che ho scritto per V è quella giusta? La visibilità misurata è quella grezza o è già corretta con il rumore e g2?}\textcolor{blue}{Ho corretto la formula, dovrebbe esserci un per due. La visibilità è quella grezza quindi credo vada bene così.}

We repeat the same interference scheme with VV states such as $\ket{\Phi^+}$ and $\ket{\Phi^-}$ (see Fig. \ref{fig:intraparticle}). For these classes of states we note that they are symmetric with respect to the BS operation.
%the polarization patterns always involved linear polarizations (see Fig. \ref{fig:intraparticle}).
This means that the reflected photon and the transmitted one always display the same state if they are indistinguishable at the input faces of the beam-splitter (see Supplementary information).
The resulting HOM correlations for the input state $\ket{\Phi^+}_a\ket{\Phi^-}_b$ are reported in  Fig. \ref{fig:HOM}-c and the achieved visibility  is equal to $V_{\ket{\Phi^+},\ket{\Phi^-}}=0.70 \pm 0.10 \%$, as expected. On the contrary, %incoming with horizontal polarization on both the input q-plates, indistinguishable
when the two photons are prepared in the same VV states such as $\ket{\Phi^+}_a\ket{\Phi^+}_b$, % with $\psi=0$ are generated and 
the theoretical HOM visibility is 1. We measured $V_{\ket{\Phi^+},\ket{\Phi^+}}=88.2 \pm 0.3 \%$, as reported in Fig. \ref{fig:HOM}-d. 

A further peculiar configuration is when the interfering input states are neither equal nor orthogonal, for which we expect a $V=\frac{1}{2}$. This is the case of two photons prepared in the input ports as $\ket{\Phi^+}_a\ket{R,2}_b $. The measured visibility is %$V_{\ket{\Phi^+},\ket{R,2}}=44.45 \pm 0.57\%$ (Fig. \ref{fig:HOM}-e).
$V_{\ket{\Phi^+},\ket{R,2}}=44.5 \pm 0.6\%$ (Fig. \ref{fig:HOM}-e).

\subsection{2-photons quantum gate: \textit{inter}-system entanglement}
%Although for distinguishable photons all four terms in Eq. \ref{Eq:BS} subsist, by post-selecting on the two-photon coincidental detection events the resulting state contains only the terms $\ket{1,1}_{cd}$. In this condition, these are not equivalent since it is possible to discriminate which photon is transmitted or reflected by the BS. 

%Before proceeding, it is important to introduce the action of the BS on the two terms that make up the angular momentum of the individual photons. Specifically, the transmitted photon undergoes no transformation, whereas by impacting on a birefringent plane oriented at $45°$, the OAM value of the reflected photon is inverted and a complex phase $e^{i}$ is inserted between its horizontal and vertical polarization. 

%Let us consider a condition in which 
The configuration described in the previous section can be exploited to implement a multi-qubit probabilistic quantum gate  able to generate an entangled state in the hybrid space composed by OAM and polarization. %In fact, by post-selecting on the two-photon coincidence events at the BS outputs, it is possible to implement a 2-qubits probabilistic quantum gate . 
In particular, by post-selecting on the two-photon coincidence events resulting from the preparation $\ket{R,2}_a \ket{R,2}_b$, and noticing that one of the output is affected by a further reflection which introduces a phase $\pi$ between horizontal and vertical polarization, and inverts the OAM value, the following maximally entangled state is generated:

%setting the orientation of the waveplates to enter the input q-plate with circular polarization  $\ket{R}$, the resulting output state result to be that one reported in Eq. \ref{Eq:BS_OAM}. However, by post-selecting on the two-photon coincidence events and noting that one of the output is affected by a further reflection which introduce a phase $\pi$ between horizontal and vertical polarization, and invert the OAM value, the following maximally entangled state is generated:

\begin{equation}
\begin{split}
    \ket{\Phi}&=\frac{\ket{L,-2}_c \ket{R,2}_d +\ket{R,2}_c \ket{L,-2}_d}{\sqrt{2}}
    =\\
    &=\frac{\ket{1,0}\ket{0,1}+\ket{0,1} \ket{1,0}}{\sqrt{2}},
\end{split}
\label{eq:gate_output}
\end{equation}

where we took off the direction subscript $\{c,d\}$ and we identified $\ket{L,-2}=\ket{1,0}$ and $\ket{R,2}=\ket{0,1}$.
Therefore, the generated state is a maximally entangled state in the 4-dimensional OAM-SAM Hilbert space. However, in the hybrid OAM-SAM space, this state can be also considered equivalent to a two-dimensional maximally entangled state. Indeed, relabeling the state $\ket{L,-2}$ as qubit $\ket{0}$ and the state $\ket{R,2}$ as qubit $\ket{1}$, the state in Eq. \eqref{eq:gate_output} results to be equivalent to a triplet Bell state:

\begin{equation}
\begin{split}
    \ket{\Phi}&=\frac{\ket{L,-2}_c \ket{R,2}_d +\ket{R,2}_c \ket{L,-2}_d}{\sqrt{2}}
    =\\
    &=\frac{\ket{0}\ket{1}+\ket{1} \ket{0}}{\sqrt{2}}
\end{split}
\label{eq:gate_output_2}
\end{equation}

Therefore, this state exhibit quantum correlations that could be detected by performing a Bell-like test which is used as an entanglement witness. In particular, we evaluated a CHSH-like inequality performing the projective measurements placing the OAM measurement stage, reported in Fig. \ref{fig:Exp_setup}, on both BS outputs. Collecting data for $400$ s and with a coincidence rate of $146$ Hz, we obtained a raw violation of $S^{(raw)} = 2.516 \pm 0.006$ which exceeds the classical bound by $86$ standard deviations, while the value obtained by subtracting the accidental coincidences is $S = 2.779 \pm 0.006$ which exceeds the separable bound by $130$ standard deviations. 

\begin{figure}[ht!]
    \centering
    \includegraphics[width=\columnwidth]{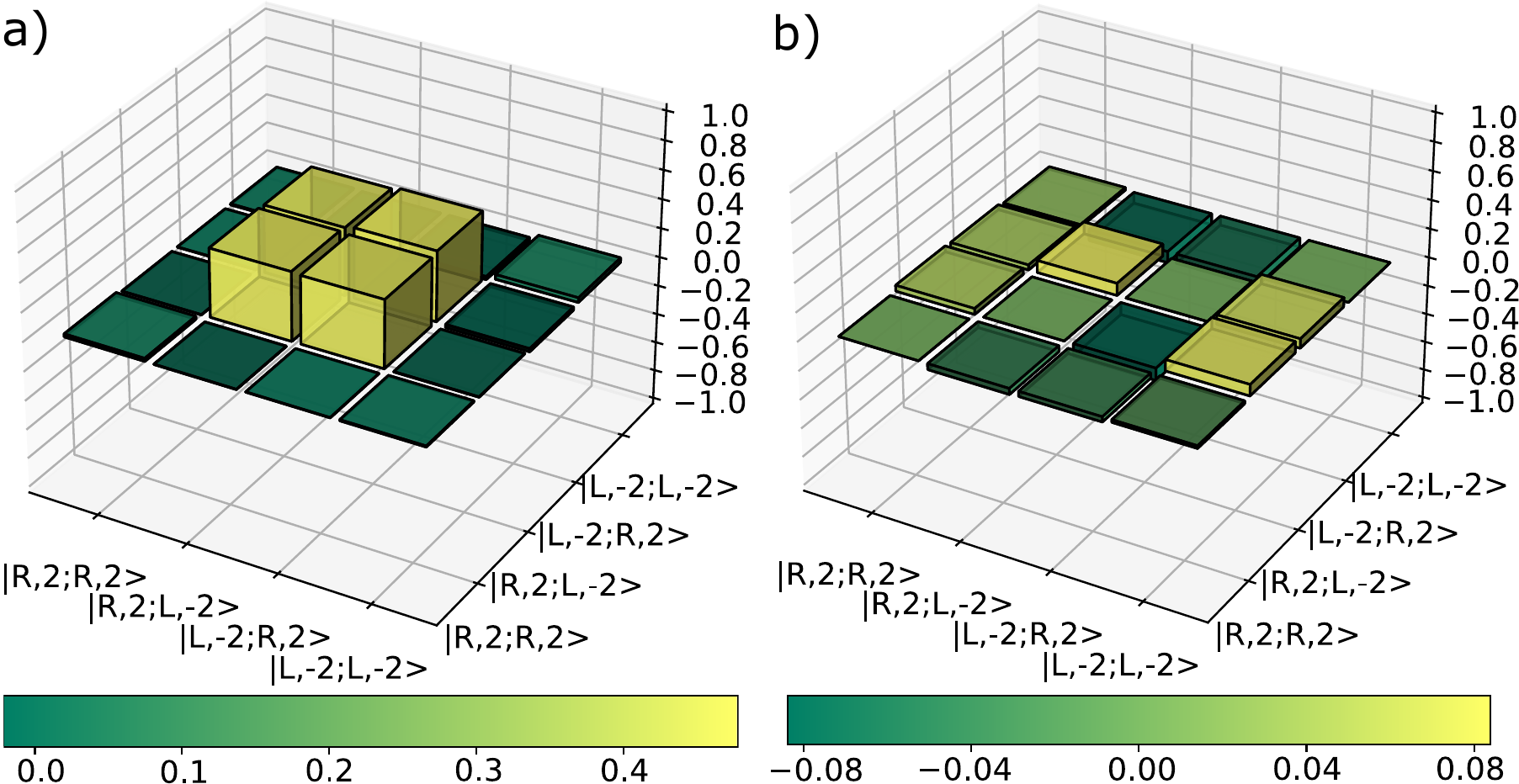}
    \caption{\textbf{\textit{Inter}-particle entangled state:} Real (a) and imaginary (b) parts of the measured density matrix for the two photons state in the hybrid OAM-polarization space reported in Eq. \eqref{eq:gate_output_2}, these are reconstructed via quantum state tomography. The fidelity between the reconstructed state and the theoretical one is equal to $\mathcal{F}=0.935 \pm 0.002$, where the standard deviation are estimated through a Monte Carlo approach assuming a Poissonian statistics.}
    \label{fig:interparticle}
\end{figure}

Moreover, we also performed a complete quantum state tomography of the state using the same experimental configuration. The retrieved density matrix is shown in Fig. \ref{fig:interparticle}, analyzing the fidelity with the triplet Bell state, we obtained a value of $\mathcal{F}=0.935 \pm 0.002$ by subtracting for accidental coincidences. The results are summarized in table \ref{Tab:Fid}. It is worth noting that the decrease in the coincidence rate is mainly due to the coupling efficiency into SMFs in the detection stage of about $ 45\%$ (see Supplemental Information for further details). This lower value depends on both the limited conversion efficiency of the QPs and on the higher divergence to which beams endowed with orbital angular momentum are subjected. Looking toward gates with more than two photons, the rate could be improved by compensating for losses due to the divergence.

\begingroup
\begin{table}[htp]
\centering
\begin{tabular}{ 
  c
  c
  c
  c
  c
  c}
\toprule
State & T (s) & Rate (Hz) & $S^{(row)}$ & $S$ & $\mathcal{F}$ \\
\midrule
\;&\;&\;\\
Intra &  $20$ & $99000$ & $2.736(8)$ & $2.792(8)$ & $0.9714(7)$ \\
\;&\;&\;\\
Inter &  $400$ &$146$ & $2.516(6)$ & $2.779(6)$ & $0.935(2)$ \\
\;&\;&\;\\
\bottomrule
\end{tabular}
\caption{\textbf{Experimental Results:} The table shows the results obtained both for the \textit{intra}-particle and \textit{inter}-particle regime. Here are reported the measurement acquisition time T, the generation rate and the values for the Bell parameter ($S$) and the fidelity. In particular, the violation $S^{(raw)}$ is computed using raw data, while the parameter $S$ is obtained subtracting the background signal or the  accidental coincidence, respectively. The fidelity value is computed comparing the reconstructed density matrix with the triplet Bell state. }
\label{Tab:Fid}
\end{table}
\endgroup

\section{Conclusions}
\label{sec:5}
In this paper, we experimentally implemented a platform capable of generating on demand photonic quantum states in high-dimensional Hilbert spaces. This was achieved by combining a bright QD source with q-plates, devices capable of coupling OAM and polarization of single photons, placed in an interferometric configuration. After assessing the properties of the source, such as the multiphoton component, and the indistinguishability of the emitted photons, we focused on the generation and analysis of entangled states in the hybrid space composed of orbital angular momentum and polarization.
The setup allows us to study both the \textit{intra}- and \textit{inter}-particle entanglement. For the former, we generated a VV state using only the engineering stage placed in one arm of the interferometer, while for the latter we exploited the interference between modulated single photons generated by the QD in two consecutive excitations to implement a probabilistic quantum gate capable of producing entangled two-photons states. The characterization of the interferometer scheme was preliminarily performed by evaluating the overlap between quantum states of single photons encoded in the hybrid Hilbert space. In particular, we observed high HOM visibilities for single photons that turn out to be indistinguishable in the detection stage, while very low visibility was observed for orthogonal quantum states. The qualities of both \textit{intra}- and \textit{inter}- particle hybrid entangled states were evaluated by performing quantum state tomography and by using Bell tests to estimate the CHSH inequality. The high values of fidelities and inequality violations highlights the performances of the proposed setup for the engineering of high-dimensional entangled states. 

In summary, we proposed and implemented experimentally a flexible platform able to generate both nearly-deterministic single-photon states that exhibit entanglement between OAM and SAM degrees of freedom, and two-photon entangled states in an Hilbert space with dimensions up to four. The employed simple and effective scheme could be extended to the multi-photon regime, opening the way to high-dimensional multi-photon experiments, whose scalability is extremely demanding for platforms based on probabilistic sources. In conclusion, the results demonstrated in the present manuscript can provide advances both for fundamental investigations and quantum photonic applications.

\section*{ACKNOWLEDGEMENTS}
This work is supported by the European Union’s Horizon 2020 research and innovation programme under the PHOQUSING project GA no. 899544, by the European Union’s Horizon 2020 Research and Innovation Programme QUDOT-TECH under the Marie Sklodowska-Curie Grant Agreement No. 86109.

%\bibliography{vvb.bib}

\end{document}

\title{Orbital angular momentum based \textit{intra}- and \textit{inter}- particle entangled states generated via a quantum dot source}

\author{Alessia Suprano}

\author{Danilo Zia}
\address{Dipartimento di Fisica, Sapienza Universit\`{a} di Roma, Piazzale Aldo Moro 5, I-00185 Roma, Italy}

\author{ Mathias Pont} 
\address{Centre for Nanosciences and Nanotechnology, CNRS, Universit\'e Paris-Saclay,
UMR 9001,10 Boulevard Thomas Gobert, 91120, Palaiseau, France}

\author{Taira Giordani}

\author{Giovanni Rodari }
\address{Dipartimento di Fisica, Sapienza Universit\`{a} di Roma, Piazzale Aldo Moro 5, I-00185 Roma, Italy}

\author{Mauro Valeri}
\address{Dipartimento di Fisica, Sapienza Universit\`{a} di Roma, Piazzale Aldo Moro 5, I-00185 Roma, Italy}

\author{Bruno Piccirillo }
\address{Dipartimento di Fisica "Ettore Pancini", Universit\`{a} Federico II, Complesso Universitario di Monte Sant'Angelo, Via Cintia, 80126 Napoli, Italy}
\address{INFN – Sezione di Napoli, Via Cintia, 80126 Napoli, Italy}

\author{Gonzalo Carvacho }
\address{Dipartimento di Fisica, Sapienza Universit\`{a} di Roma, Piazzale Aldo Moro 5, I-00185 Roma, Italy}

\author{Nicolò Spagnolo }
\address{Dipartimento di Fisica, Sapienza Universit\`{a} di Roma, Piazzale Aldo Moro 5, I-00185 Roma, Italy}

\author{Pascale Senellart}
\address{Centre for Nanosciences and Nanotechnology, CNRS, Universit\'e Paris-Saclay,
UMR 9001,10 Boulevard Thomas Gobert, 91120, Palaiseau, France}

\author{Lorenzo Marrucci}
\address{Dipartimento di Fisica "Ettore Pancini", Universit\`{a} Federico II, Complesso Universitario di Monte Sant'Angelo, Via Cintia, 80126 Napoli, Italy}

\author{Fabio Sciarrino }

\email{fabio.sciarrino@uniroma1.it}

\address{Dipartimento di Fisica, Sapienza Universit\`{a} di Roma, Piazzale Aldo Moro 5, I-00185 Roma, Italy}

\maketitle
\newpage
\section{Theoretical background}
In this section, we provide a theoretical insight about the different HOM interference cases reported in the main text. Recalling that the action of a symmetric
Beam Splitter (BS) with input modes $\{a,b\}$ and output modes $\{c,d\}$ can be expressed as:

\begin{equation}
\begin{pmatrix}
\hat{a} \\
\hat{b}
\end{pmatrix}=\frac{1}{\sqrt{2}}
\begin{pmatrix}
1 & 1\\
1 &-1
\end{pmatrix}
\begin{pmatrix}
\hat{c} \\
\hat{d}
\end{pmatrix}
\end{equation}
Therefore, by considering two photons at the two inputs of the beam splitter, the initial state is $\hat{a}^{\dagger} \hat{b}^{\dagger} \ket{0,0}$ and the output state results to be:
%associate two bi-dimensional Hilbert spaces $\mathcal{H}_a$ and $\mathcal{H}_b$ to the pair of single photons incident on a 50:50 beam splitter, where the basis elements $\{\ket{0}, \ket{1}\}$ are associated to the two inputs. The action of the bulk-BS, can be schematize with the 2-qubits quantum gate reported in Fig. \ref{}. Therefore, applying this gate to the initial state $\ket{0}_a \ket{1}_b$, the output state can be expressed as the combination of 4 distinct terms:
\begin{equation}\label{Eq:BS}
%\begin{aligned}   \frac{1}{2}(\hat{c}^{\dagger}-\hat{d}^{\dagger})(\hat{c}^{\dagger}+\hat{d}^{\dagger})\ket{0,0}_{cd}=\\\frac{1}{2}(\ket{2,0}_{cd}+\ket{1,1}_{cd}-\ket{1,1}_{cd}-\ket{0,2}_{cd})%\end{aligned}
\frac{1}{2}\left( \hat{c}^{\dagger}_a \hat{c}^{\dagger}_b + \hat{c}^{\dagger}_a \hat{d}^{\dagger}_b - \hat{c}^{\dagger}_b \hat{d}^{\dagger}_a - \hat{d}^{\dagger}_a \hat{d}^{\dagger}_b\right) \ket{0,0},
\end{equation}
%where the terms $\ket{2,0}_{cd}$ and $\ket{0,2}_{cd}$ are associated to the probability with which the photons come out from the same output, while $\ket{1,1}_{cd}$ and $\ket{1,1}_{cd}$ represent the case in which the photons come out of different outputs. 
where $\ket{0,0}$ refers to the vacuum in the two ports of the BS and the subscripts $a$ and $b$ to the field modes at the inputs. %\commDZ{Qui credete vada fatta un'immagine che mostra il BS e i modi d'ingresso e di uscita? Lmk, secondo me ptrebbe essere utile}
%Since single photons are characterized by different degrees of freedom such as polarization, OAM and time-frequency, their state depends on each of them and thus it is important to discriminate between distinguishable and undistinguishable photons.
Two photons are indistinguishable if their state, associated to each degree of freedom, is the same from the point of view of the observer. In this case, when the two initial modes $a$ and $b$ are identical, it is not possible for the observer to discriminate which one of the two photons come out from the outputs $c$ or $d$. In other words, the term $\hat{c}^{\dagger}_a \hat{d}^{\dagger}_b - \hat{c}^{\dagger}_b \hat{d}^{\dagger}_a$ vanishes.

For what concern the BS action on the polarization and Orbital Angular Momentum (OAM) states, we have that the horizontal and vertical polarizations result to be eigenstates of the BS operation since the reflection introduces a phase $\phi=\pi$ between them. On the other side, in the OAM space, the eigestates are the balanced superpositions $\frac{\ket{m}+\ket{-m}}{\sqrt{2}}$ and  $\frac{\ket{m}-\ket{-m}}{\sqrt{2}}$ which are not affected by the elicity filp.\\
For instance, considering the case in which the waveplates change the input photons polarization to $\ket{R}$ in both of the arms, after passing through the q-plate (see Main Text, Eq. 1), at the exit of the BS we have:\\
\begin{equation}\label{Eq:BS_OAM}
\begin{aligned}
    \frac{1}{2}(\hat{c}^{\dagger}_{L,-2}-\hat{d}^{\dagger}_{R,2})(\hat{c}^{\dagger}_{R,2}+\hat{d}^{\dagger}_{L,-2})\ket{0,0}_{cd}=\\
    =\frac{1}{2}(\ket{L,-2}_{c}\ket{R,2}_{c}+\ket{L,-2}_{c}\ket{L,-2}_{d}-\\\ket{R,2}_{c}\ket{R,2}_{d}-\ket{R,2}_{d}\ket{L,-2}_{d})
\end{aligned}
\end{equation}
In this case is not possible to extinguish the double coincidence terms and to see the HOM interference we need to enter on the q-plates with orthogonal polarization states. For example, with $\ket{R}_{a}$ and $\ket{L}_{b}$ we have:
\begin{equation}\label{Eq:BS_OAM_3}
\begin{aligned}
    \frac{1}{2}(\hat{c}^{\dagger}_{L,-2}-\hat{d}^{\dagger}_{R,2})(\hat{c}^{\dagger}_{L,-2}+\hat{d}^{\dagger}_{R,2})\ket{0,0}_{cd}=\\
    %=\frac{1}{2}(\sqrt{2}\ket{L,-2}_{c}\ket{L,-2}_{c}+\ket{L,-2}_{c}\ket{R,2}_{d}-\\\ket{L,-2}_{c}\ket{R,2}_{d}-\sqrt{2}\ket{R,2}_{d}\ket{R,2}_{d})=\\
=\frac{1}{\sqrt{2}}(\ket{L,-2}_{c}\ket{L,-2}_{c}-\ket{R,2}_{d}\ket{R,2}_{d})
\end{aligned}
\end{equation}

This behaviour is experimentally observed in our setup as reported in Fig. 5 (a,b) of the Main Text.
If instead we enter the q-plates with linearly polarized photons on both the arms, at the output of the BS we observed the complementary behavior in which we observed coincidences only when the photons have orthogonal polarization. In fact considering the input states $\ket{H}_{a}\ket{H}_{b}$ and $\ket{H}_{a}\ket{V}_{b}$, after the q-plates we have $\ket{\Phi^+}_{a}\ket{\Phi^+}_{b}$ and$\ket{\Phi^+}_{a}\ket{\Phi^-}_{b}$, then the BS output is respectively:\\
\begin{equation}
\begin{aligned}\label{eq:BS_HH}
    a^{\dagger}_{\Phi^+}b^{\dagger}_{\Phi^+} \longmapsto & \frac{1}{4}(\hat{c}^{\dagger}_{R,2}-\hat{d}^{\dagger}_{L,-2}+\hat{c}^{\dagger}_{L,-2}-\hat{d}^{\dagger}_{R,2})(\hat{c}^{\dagger}_{L,-2}+\hat{d}^{\dagger}_{R,2}+\hat{c}^{\dagger}_{R,2}+\hat{d}^{\dagger}_{L,-2})\ket{0,0}_{cd}=\\
    %=\frac{1}{2}(\ket{L,-2}_{c}\ket{VVB}_{c}+\ket{L,-2}_{c}\ket{VVB}_{d}-\\\ket{VVB}_{c}\ket{R,2}_{d}-\ket{VVB}_{d}\ket{R,2}_{d})=\\
&=\frac{1}{4}(\sqrt{2}\ket{L,-2}_{c}\ket{L,-2}_{c}+\sqrt{2}\ket{R,2}_{c}\ket{R,2}_{c}+2\ket{L,-2}_{c}\ket{R,2}_{c}+\\&-\sqrt{2}\ket{L,-2}_{d}\ket{L,-2}_{d}-2\ket{L,-2}_{d}\ket{R,2}_{d}-\sqrt{2}\ket{R,2}_{d}\ket{R,2}_{d})
\end{aligned}
\end{equation}
\\
\begin{equation}
\begin{aligned}\label{eq:BS_HV}
    a^{\dagger}_{\Phi^+}b^{\dagger}_{\Phi^-} \longmapsto & \frac{1}{4}(\hat{c}^{\dagger}_{R,2}-\hat{d}^{\dagger}_{L,-2}+\hat{c}^{\dagger}_{L,-2}-\hat{d}^{\dagger}_{R,2})(\hat{c}^{\dagger}_{L,-2}+\hat{d}^{\dagger}_{R,2}-\hat{c}^{\dagger}_{R,2}-\hat{d}^{\dagger}_{L,-2})\ket{0,0}_{cd}=\\
    %=\frac{1}{2}(\ket{L,-2}_{c}\ket{VVB}_{c}+\ket{L,-2}_{c}\ket{VVB}_{d}-\\\ket{VVB}_{c}\ket{R,2}_{d}-\ket{VVB}_{d}\ket{R,2}_{d})=\\
&=\frac{1}{4}(\sqrt{2}\ket{L,-2}_{c}\ket{L,-2}_{c}-\sqrt{2}\ket{R,2}_{c}\ket{R,2}_{c}+2\ket{L,-2}_{c}\ket{L,-2}_{d}+\\&+\sqrt{2}\ket{L,-2}_{d}\ket{L,-2}_{d}+2\ket{R,2}_{c}\ket{R,2}_{d}
-\sqrt{2}\ket{R,2}_{d}\ket{R,2}_{d})
\end{aligned}
\end{equation}
According to Eq. \eqref{eq:BS_HH} no coincidences are expected, while in Eq. \eqref{eq:BS_HV} the terms in which each photon exit to different path do not cancel out and we have a coincidences probability of $p_{c,d}=\frac{1}{2}$. The expected visibility is then $V^{th}_{\Phi^+\Phi^+}=1-2p_{c,d}=1$ and $V^{th}_{\Phi^+\Phi^-}=0$, these are in agreement with the measured behaviours reported in Fig. 5 (c,d).\\
Moreover, we analyze also the case  when interfering input states are neither equal nor orthogonal. Let us consider to income on the input q-plates with polarization $|R\rangle_a$ and $|H\rangle_b$. In the first arm the generated state is $|L,-2\rangle_a$, while a VV state $\ket{\Phi^+}_{b}$ is generated in the second arm. Tacking into account the BS transformation, the output beam result to be:
\begin{equation}\label{Eq:BS_OAM_2}
\begin{aligned}
    \frac{1}{2}(\hat{c}^{\dagger}_{L,-2}-\hat{d}^{\dagger}_{R,2})(\hat{c}^{\dagger}_{R,2}+\hat{d}^{\dagger}_{L,-2}+\hat{c}^{\dagger}_{L,-2}+\hat{d}^{\dagger}_{R,2})\ket{0,0}_{cd}=\\
    %=\frac{1}{2}(\ket{L,-2}_{c}\ket{VVB}_{c}+\ket{L,-2}_{c}\ket{VVB}_{d}-\\\ket{VVB}_{c}\ket{R,2}_{d}-\ket{VVB}_{d}\ket{R,2}_{d})=\\
=\frac{1}{2\sqrt{2}}(\sqrt{2}\ket{L,-2}_{c}\ket{L,-2}_{c}+\ket{L,-2}_{c}\ket{R,2}_{c}+\\
+\ket{L,-2}_{c}\ket{L,-2}_{d}-
\ket{R,2}_{c}\ket{R,2}_{d}-\\
-\ket{R,2}_{d}\ket{L,-2}_{d}-\sqrt{2}\ket{R,2}_{d}\ket{R,2}_{d})
\end{aligned}
\end{equation}
%Where the $\sqrt{2}$ is related to the creation operator's action when the Fock number state is different from the void, i. e. $a^{\dagger}\ket{n}=\sqrt{n+1}\ket{n+1}$. This happen only when all the degrees of freedom are the same. 
In this condition the probability of having both photons on the same output is $p_{c,c}+p_{d,d}=\frac{3}{4}$ %\frac{2}{3}$, 
while the probability of having  the photons come out of different outputs is $p_{c,d}=\frac{1}{4}$. %\frac{1}{3}$. 
Therefore, we expect an HOM visibility of $V_{R,H}^{Th}=\frac{1}{2}$%\frac{1}{3}$
, which is consistent with the experimental one reported in the Main Text.

\section{Efficiency estimation}

%In this section, we report a step-by-step estimation of the experimental platform losses. The QD source is pumped with a $79$~MHz-pulsed laser and it generates deterministically one photon at each pulse ($R_{\mathrm{exe}}=79$ MHz). However, the single photons detected with the Avalanche Photo-Diode detector characterized by an efficiency of $\eta_{\mathrm{det}}\sim 38\%$ at the output of the \textit{e-Delight} system are $3.8$ MHz, therefore as output of the source we have $10$ MHz with an overall efficiency of $\eta_{\mathrm{source}}\sim 12.6\%$. Subsequently, the photons are transported from the source to the implemented apparatuses through optical fibers with transmission efficiency of connected by mating sleeves, each of which has an overall efficiency of $\eta_{\mathrm{fiber}}\sim 80\%$. On the experimental platform, the photons with horizontal polarization are selected through a PBS and, since there is a residual of vertical polarized photons generated by the QD, the PBS has a transmittivity of $\eta_{\mathrm{PBS}}\sim 83\%$. Then, the major sources of losses are the q-plates which have an efficiency of $\eta_{\mathrm{q-plate}}\sim 70\%$  and the coupling with $\eta_{\mathrm{coupling}}\sim 45\%$. %Moreover, we used Avalanche Photo-Diode detectors whose efficiency at $930$ nm is around $\eta_{\mathrm{det}}\sim 38\%$.

In this section, the transmission of all optical components are reported to estimate the efficiency of the experimental platform. The single-photon countrate $C^{te}=4$~MHz is measured at the output of the Q-Fiber (Quandela) module with an Avalanche Photo-Diode (APD) detector with $\eta_{\mathrm{det}}\sim 38\%$ efficiency. Knowing the QD source is pumped with a $79$~MHz-pulsed laser ($R_{\mathrm{exe}}=79$ MHz) and accounting for the limited efficiency of the APD, we can derive the fibered brightness of the single-photon source $\eta_{\mathrm{fibered}}\sim 13.3\%$. The Q-Fiber module is connected to the experimental platform through optical fibers. Each connection is made with FC/PC mating sleeves, which each have a transmission of $\eta_{\mathrm{connector}}\sim 80\%$, while the fiber-BS employed in the \textit{inter}-particle regime has an insertion loss of $\eta_{\mathrm{BS}}\sim 0.75$. The single-photon are then filtered in polarization with polarizing beamsplitters (PBS) and, since there is a residual of vertical polarized photons generated by the QD, the efficiency is $\eta_{\mathrm{pol}}\sim 83\%$. The major sources of losses are the q-plates which have an efficiency of $\eta_{\mathrm{q-plate}}\sim 70\%$ and the coupling of a VV beam converted to gaussian mode with a SMF $\eta_{\mathrm{coupling}}\sim 45\%$.
In conclusion, the overall efficiency of the generation stage is $\eta_{\mathrm{gen,1}}=\eta_{\mathrm{fibered}} \eta _{\mathrm{connector}}^2 \eta_{\mathrm{pol}} \eta_{\mathrm{q-plate}}=4.95 \%$ and $\eta_{\mathrm{gen,2}}=\eta_{\mathrm{fibered}} \eta _{\mathrm{connector}}^3 \eta_{\mathrm{BS}}\eta_{\mathrm{pol}} \eta_{\mathrm{q-plate}}=2.97 \%$ respectively for \textit{intra}- and \textit{inter}-particle regime, while the tomography setup results to have an efficiency of $\eta_{\mathrm{tomo}}=\eta_{\mathrm{q-plate}} \eta_{\mathrm{coupling}}\eta_{\mathrm{det}}=11.97 \%$.
%the apparatuses and the single photon source is $\eta=\eta_{\mathrm{source}}\eta_{\mathrm{sleeves}}^2\eta_{\mathrm{q-plate}}^2\eta_{\mathrm{coupling}}\eta_{\mathrm{det}}=0.37 \%$. 
Using such approximated efficiency, it is possible to estimate the expected signals in the different configurations of the experiment. For the \textit{intra}-particle entanglement, all the measurements are performed on single events, and by taking into account the $50\%$ loss due to the second beamsplitter (BS) of the interferometer, the expected generation rate is $R^{gen}_{\mathrm{intra}}=\frac{\eta_{\mathrm{gen,1}} R_{\mathrm{exe}}}{2}=1.96$ MHz, while the detected countrate should be $R_{\mathrm{intra}}=\eta_{tomo} R^{gen}_{\mathrm{intra}}=234.1$ kHz. On the other side, in the \textit{inter}-particle experiment, the measurements are performed detecting the coincidence counts, therefore the expected generation rate is $R^{gen}_{\mathrm{inter}}=\frac{\eta_{\mathrm{gen,2}}^2 R_{\mathrm{exe}}}{4 \cdot 2}=8.71$ kHz while the detected coincidence rate is $R^{gen}_{\mathrm{inter}}=\eta_{\mathrm{tomo}}^2 R^{gen}_{\mathrm{inter}}=124.8$ Hz, where the coefficients $\frac{1}{4}$ and $\frac{1}{2}$ are due to the first and second BS of the interferometer, respectively.
%The estimated coounts in both configurations are very close to the measured ones $R_{\mathrm{intra}}^{\mathrm{exp}}=99$ kHz and $R_{\mathrm{inter}}^{\mathrm{exp}}=146$ Hz.

\begingroup
\setlength{\tabcolsep}{1 pt}% Default value: 6pt
\begin{table}[htp]
\centering
\begin{tabular}{ 
  c|
  c|
  c|
  c|
  c|
  c|
  c|
  c|
  c|
  c|
  c}
\toprule
$\eta_{\mathrm{fibered}}$&
$\eta_{\mathrm{connector}}$&
$\eta_{\mathrm{BS}}$&
$\eta_{\mathrm{pol}}$&
$\eta_{\mathrm{q-plate}}$ & $\eta_{\mathrm{coupling}}$ & $\eta_{\mathrm{det}}$ & $R^{gen}_{\mathrm{intra}}$ (MHz) & $R_{\mathrm{intra}}$ (kHz)& $R^{gen}_{\mathrm{inter}}$ (kHz) & $R_{\mathrm{inter}}$ (Hz)\\
\midrule

$13.3\%$ & $80\%$ & $75\%$& $83\%$ & $70\%$ & $45\%$ & $38\%$ & $1.96$ & $234.1$ & $8.71$ & $124.8$\\

\bottomrule
\end{tabular}
\caption{\textbf{Efficiency estimation:} The table showcases the experimental efficiency estimated for each loss of the apparatus. 
$\eta_{\mathrm{fibered}}$ represents the overall transmission efficiency of the single-photon source, including excitation laser rejection and extraction system losses. $\eta_{\mathrm{connector}}$ is the transmission efficiency of single-mode optical fiber, $\eta_{\mathrm{BS}}$ the insertion loss of the fiber-BS, while $\eta_{\mathrm{q-plate}}$ and $\eta_{\mathrm{coupling}}$ are the main losses of the implemented experimental platform due to the limited conversion efficiency of OAM modes and the higher divergence to which beams endowed with orbital angular momentum are subjected. Finally, $\eta_{\mathrm{det}}$ is the intrinsic efficiency of the Avalanche Photo-Diode detectors employed. Using these values the expected generated and detected count rates are evaluated for both the \textit{intra}- ($R^{gen}_{\mathrm{intra}}$ and $R_{\mathrm{intra}}$) and \textit{inter}-particle ($R^{gen}_{\mathrm{inter}}$ and $R_{\mathrm{inter}}$) regime. }
\label{Tab:Fid}
\end{table}
\endgroup